\documentclass{aa}  
\usepackage{graphicx}
\usepackage{txfonts}
\usepackage{natbib}
\usepackage{hyperref}
\hypersetup{colorlinks=True,citecolor=blue,allcolors=blue}
\usepackage{tabu}
\hypersetup{draft}

\begin{document} 

\title{Revealing the structure of the lensed quasar Q 0957+561}
\subtitle{I. Accretion disk size}

\author{C. Fian\inst{1}, E. Mediavilla\inst{2,3}, J. Jim\'enez-Vicente\inst{4,5}, V. Motta\inst{6}, J. A. Mu\~noz\inst{7,8}, D. Chelouche\inst{9}, P. Gom\'ez-Alvarez\inst{10}, K. Rojas\inst{11}, A. Hanslmeier\inst{12}}
\institute{School of Physics and Astronomy and Wise Observatory, Raymond and Beverly Sackler Faculty of Exact Sciences, Tel-Aviv University, Tel-Aviv, Israel \and Instituto de Astrof\'{\i}sica de Canarias, V\'{\i}a L\'actea S/N, La Laguna 38200, Tenerife, Spain \and Departamento de Astrof\'{\i}sica, Universidad de la Laguna, La Laguna 38200, Tenerife, Spain \and Departamento de F\'{\i}sica Te\'orica y del Cosmos, Universidad de Granada, Campus de Fuentenueva, 18071 Granada, Spain \and Instituto Carlos I de F\'{\i}sica Te\'orica y Computacional, Universidad de Granada, 18071 Granada, Spain \and Instituto de F\'{\i}sica y Astronom\'{\i}a, Universidad de Valpara\'{\i}so, Avda. Gran Breta\~na 1111, Playa Ancha, Valpara\'{\i}so 2360102, Chile \and Departamento de Astronom\'{i}a y Astrof\'{i}sica, Universidad de Valencia, E-46100 Burjassot, Valencia, Spain \and Observatorio Astron\'{o}mico, Universidad de Valencia, E-46980 Paterna, Valencia, Spain \and Haifa Research Center for Theoretical Physics and Astrophysics, University of Haifa, Haifa, Israel \and FRACTAL S.L.N.E., Calle Tulip\'an 2, Portal 13, 1A, E-28231 Las Rozas de Madrid, Spain \and Institute of Physics, Laboratoire d'Astrophysique, Ecole Polytechnique Fédérale de Lausanne (EPFL), Observatoire de Sauverny, CH-1290 Versoix, Switzerland \and Institute of Physics (IGAM), University of Graz, Universit{\"a}tsplatz 5, 8010, Graz, Austria}
%\date{Received xxxxxxx; accepted xxxxxxx}
\abstract
{}
{We aim to use signatures of microlensing induced by stars in the foreground lens galaxy to infer the size of the accretion disk in the gravitationally lensed quasar Q 0957+561. The long-term photometric monitoring of this system (which so far has provided the longest available light curves of a gravitational lens system) permits us to evaluate the impact of uncertainties on our recently developed method (controlled by the distance between the modeled and the experimental magnitude difference histograms between two lensed images), and thus to test the robustness of microlensing-based disk-size estimates.}
{We analyzed the well-sampled 21-year GLENDAMA optical light curves of the double-lensed quasar and studied the intrinsic and extrinsic continuum variations. Using accurate measurements for the time delay between the images A and B, we modeled and removed the intrinsic quasar variability, and from the statistics of microlensing magnifications we used a Bayesian method to derive the size of the region emitting the continuum at $\lambda_{rest} = 2558$\AA.}
{Analysis of the Q 0957+561 R-band light curves show a slow but systematic increase in the brightness of the B relative to the A component during the past ten years. The relatively low strength of the magnitude differences between the images indicates that the quasar has an unusually big optical accretion disk of half-light radius: $R_{1/2} = 17.6\pm6.1 \sqrt{M/0.3M_\odot}$ lt-days.}
{}


\titlerunning{Accretion disk size of Q 0957+561}
\authorrunning{Fian et al.} 
\maketitle

\section{Introduction}
The temporal changes in brightness of the images of a gravitationally lensed quasar can be described as a combination of time-delay-correlated (intrinsic variability of the source) and uncorrelated (gravitational microlensing) variations, and their analysis has important applications in cosmology such as the determination of time delays to infer the Hubble constant (\citealt{Rusu2020,Wong2020,Birrer2020}), the estimate of peculiar velocities (Mediavilla et al. 2016), and in the study of quasar structure (\citealt{Chang1979,Chang1984}; see also \citealt{Kochanek2004} and \citealt{wam2006}). Up to now it has been impossible to spatially resolve the emitting regions of quasars, even with the largest available optical telescopes. The random distribution of compact objects such as stars in the foreground lens galaxy induce uncorrelated flux anomalies in the multiple quasar images, which can help us overcome these difficulties and thus can be used to extract information about both the source and the lens itself. These (de)magnifications produced by microlenses depend strongly on the angular size of the source, with smaller emission regions showing bigger changes in the brightness, while larger sources result in smoother light curves (\citealt{MosqueraKochanek2011,Blackburne2011,Blackburne2014}).  Over the past years, quasar microlensing has therefore become a powerful tool to study the continuum emission regions of quasars by measuring and modeling the time-variable flux ratios between lensed images (e.g., \citealt{Motta2012,Blackburne2014,Blackburne2015,Jimenez2012,Jimenez2014,Mosquera2009,Mosquera2013,Mediavilla2015,Munoz2016,Fian2016,Fian2018,Morgan2018}). \\

 Q 0957+561 was discovered by Dennis Walsh in 1979. At a redshift of $z_s = 1.41$, the quasar is lensed into two bright point sources. The separation between images is $\sim$6\arcsec, and several studies (\citealt{Pelt1996,Oscoz1996,Oscoz1997,Schild1997,Kundic1997,Oscoz2001,Ovaldsen2003,Colley2003,Shalyapin2008}) reported a time delay of $\sim$14 months (see Table \ref{timedelay}) between the two images (with A being the leading component). The low microlensing variability and the lack of microlensing events in the early light curves of the first known gravitational lens Q 0957+561 prevented determinations of the quasar's accretion disk size (\citealt{Schmidt1998,Wambsganss2000}). After combining new optical monitoring data with previously published data, \citealt{Hainline2012} reported a new microlensing event and thus demonstrated the return of long-timescale, uncorrelated variability in the light curves of Q 0957+561. In total, they used $\sim$15 years of photometric monitoring (although with rather large gaps) to constrain the size of the optical accretion disk. In this work, we conducted a statistical analysis of an extended data set of 21 years of photometric monitoring, which currently features the longest available light curves of a gravitationally lensed quasar. Our aim is to use this large amount of monitoring data to study the existence of possible microlensing events and to place improved constraints on the size of the continuum-emitting region (hereafter referred to as accretion disk, although the contribution of additional continuum components may be non-negligible; \citealt{Chelouche2019}). We followed the single-epoch method combined with the flux ratios of a large enough source in the quasar in order to be insensitive to microlensing and to establish the baseline for no microlensing magnification (see, e.g., \citealt{Metcalf2002,Mediavilla2009}). The broad emission lines' cores (arising from the narrow-line region), the mid-IR (emitted by the dusty torus), and the radio emitting regions (radio jet and lobes) of quasars should all be large enough to average out the effects of microlensing and allow the determination of the "intrinsic" flux ratios between images (\citealt{Kochanek2004}). Flux ratios detected at radio wavelengths are considered as the most robust, because the sources are believed to be much larger than the Einstein radius $\eta_0$. However, radio flux ratios are measurable only for a small subsample of lensed quasars bright enough at radio wavelengths (about one out of five known systems, see \citealt{Sluse2013}), and other proxies of the unbiased intrinsic flux ratios have to be found. %A promising alternative to estimate flux ratios free of microlensing are mid-IR ratios as most of the emission is assumed to originate from the large size dusty torus (\citealt{Chiba2005,Minezaki2009}). Several authors found that the mid-IR emission region has a scale comparable to or larger than a the typical Einstein radius of a microlens (see, e.g., \citealt{Wyithe2002,Swain2003,Kishimoto2011}), and hence at least two orders of magnitude larger than the optical emission regions which is thought to be $< 0.01 \eta_0$ (\citealt{Wambsganss1990,Wyithe2000}).
 Furthermore, \citet{Guerras2013} also suggested that the cores of the broad emission lines are not subject to a large microlensing variation and can be used as a baseline for no microlensing.\\
 
 In the present work, we extended the single-epoch method to more than 1000 epochs in the available light curves, thereby increasing the statistical significance. We used the optical R-band light curves obtained from the GLENDAMA project (\citealt{Gil-Merino2018}) to infer microlensing flux variability and radio data from the literature to estimate the baseline for no microlensing variability. After correcting for the relatively long time delay and the mean magnitude difference between the images, we find clear indications of slow microlensing variability in the residuals of the light curves over the past ten years. Following the methods described in \citet{Fian2016,Fian2018}, we compared the histogram of microlensing magnifications obtained from the observations (corresponding to the monitoring time interval) with the simulated predictions of microlensing variability for sources of different sizes. This comparison allowed us to evaluate the likelihood of the different values adopted for the size. In the present study, apart from using the so far longest available light curves, we used a more rigorous method to estimate the accretion disk size, and, in addition, we evaluated the effect of different kinds of uncertainties on the size of the accretion disk.
\begin{table}[h]
\renewcommand{\arraystretch}{1.2}
\caption{Time delay measurements from the literature.}
\begin{tabu}to 0.49\textwidth {X[l]X[l]}
\hline 
\hline 
$\Delta$t (days) & Reference \\ 
\hline 
 423$\pm$6 & \citealt{Pelt1996} \\ 
 404$\pm$26 & \citealt{Schild1997} \\ 
 417$\pm$3 & \citealt{Kundic1997} \\
 422.6$\pm$0.6 & \citealt{Oscoz2001} \\
 429.9$\pm$1.2 & \citealt{Ovaldsen2003} \\
 417.09$\pm$0.07 & \citealt{Colley2003} \\
 417$\pm$2 & \citealt{Shalyapin2008} \\ \hline 
\end{tabu}
\label{timedelay}    
\end{table}

The paper is organized as follows. In Section \ref{2}, we present the full combined light curves of images A and B of Q 0957+561. We outline our modeling of the intrinsic variability and examine the flux ratios between the images in Section \ref{3}. Section \ref{4} is devoted to the Bayesian source size estimation based on the statistics of microlensing magnifications. In Section \ref{5}, we present the results and discuss the impact of uncertainties on the size estimates. Finally, we give a brief summary in Section \ref{6}.

\section{Data}\label{2}
\begin{figure} 
\centering
\includegraphics[width=8.4cm]{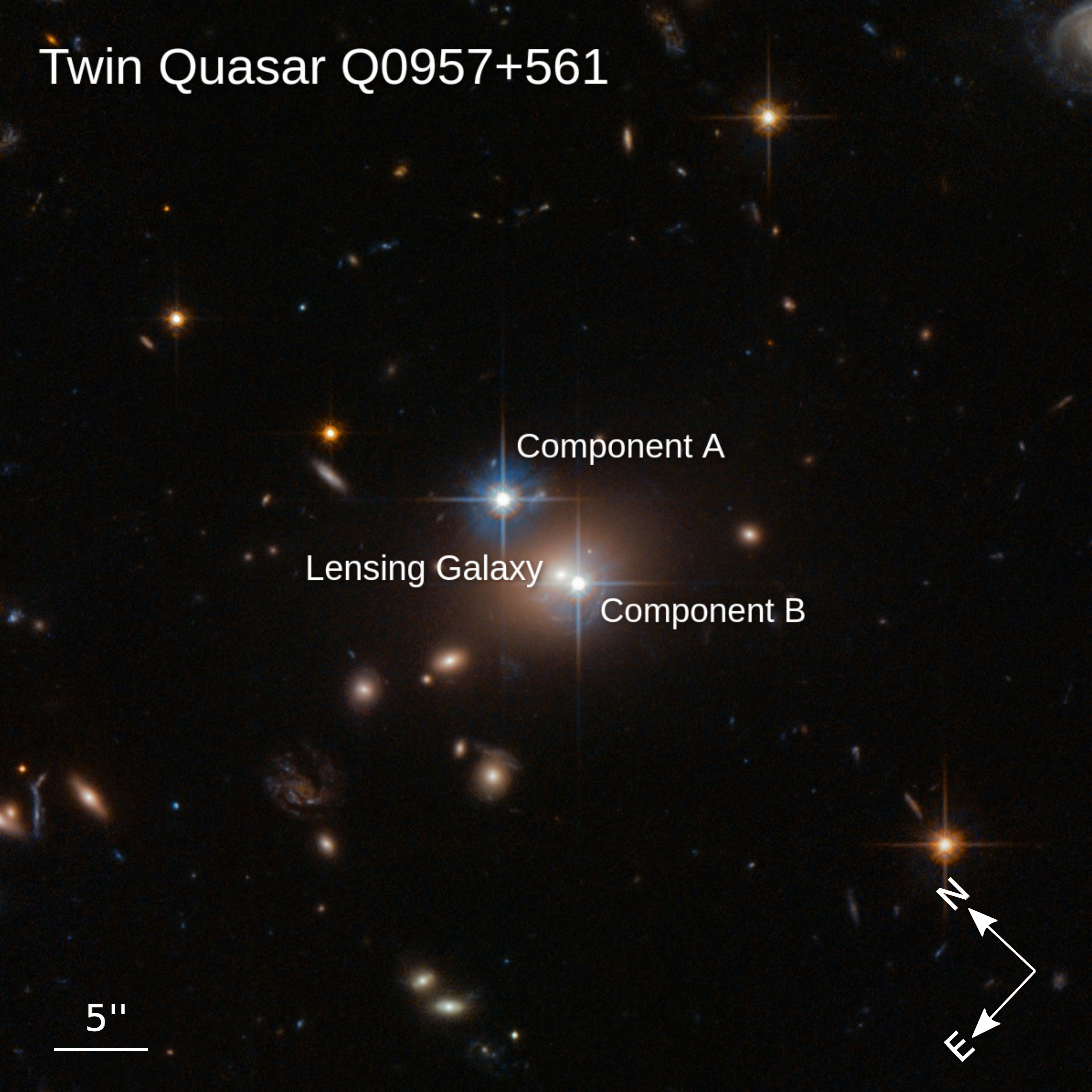}
\caption{Hubble image\protect\footnotemark\ of the gravitationally lensed quasar Q 0957+561 in the constellation Ursa Major. The two components are separated by $\sim$6\arcsec, with image B being located close ($\sim$1\arcsec) to the lensing galaxy G1, which is a giant elliptical lying within a cluster of galaxies that also contributes to the lensing. The field of view (FoV) is $\sim1$ square arcminute.}
\label{image}
\end{figure}

\begin{figure*}
\centering
\includegraphics[width=17.85cm]{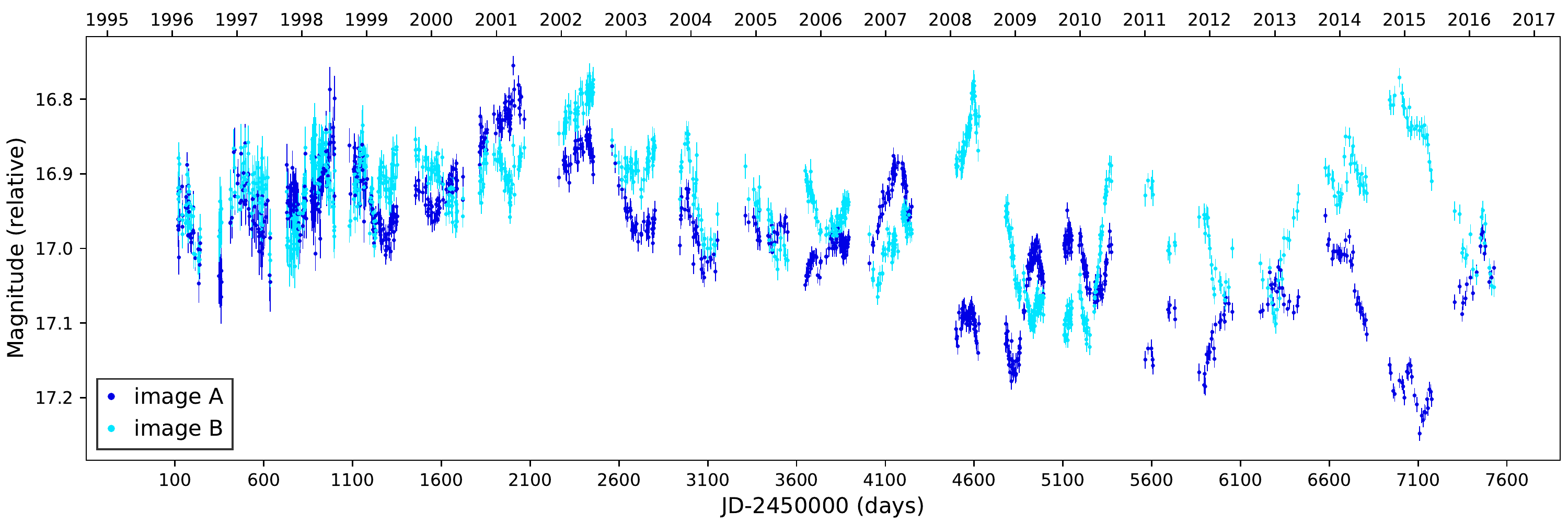}
\caption{Light curves of the lensed images A and B of the quasar Q 0957+561 from 1996 February to 2016 May as obtained by the GLENDAMA project (see \citealt{Gil-Merino2018}). Horizontal axes show the Julian (bottom) and Gregorian (upper) dates.}
\label{data}
\end{figure*}

The fluxes of the two images of Q 0957+561 were monitored from 1996 February until 2016 May in the optical R-band (at $\lambda_{rest}=2558$\AA) as a part of the Gravitational LENses and DArk MAtter (GLENDAMA) project (see \citealt{Gil-Merino2018}). They used the observations made 
at the 0.8 m telescope of the Instituto de Astrof\'isica de Canarias' (IAC) Teide Observatory (\citealt{Oscoz1996,Oscoz1997,Oscoz2001,Oscoz2002,Serra-Ricart1999}) during the first observing period (1996-2005). They later monitored the double quasar with the 2 m Liverpool Telescope (LT) at the Roque de los Muchachos Observatory from 2005 to 2016 (\citealt{Gil-Merino2018}). The data set consists of 1067 epochs of observation (i.e., 1067 nights), and the average sampling rate is once every seven days. In Table \ref{characteristics}, we list the object's characteristics. In Figure \ref{image}, an image of Q 0957+561 is shown and, in Figure \ref{data} the 21-yr light curves of the images A and B are presented.
\begin{table*}[h]
        \renewcommand{\arraystretch}{1.2}
        \caption{Q 0957+561 characteristics.}
        \begin{tabu} to \textwidth {X[c]X[c]X[c]X[c]X[c]X[c]X[c]} 

                \hline
                \hline 
                R.A. (\arcsec) & Dec. (\arcsec) & $z_s$ & $z_l$ & $N$ & $\delta\ (\arcsec)$ & r-SDSS band\\  
                (1) & (2) & (3) & (4) & (5) & (6) & (7)\\ \hline
                $0/1.229$ & $0/-6.048$ & $1.413$ & $0.3562$ & $2$ & $6.26$ & $17.5/16.9$\\ \hline 
        \end{tabu}
        \tablefoot{--- Cols. (1)--(2): Relative coordinates (right ascension and declination) of images A and B from the CASTLES Survey\footnotemark. Cols. (3)--(4): Redshift of the quasar and the lens galaxy from the Gravitationally Lensed Quasar Database\footnotemark. Cols. (5)--(6): Number of images and image separation from the CASTLES Survey. Col. (7): r-SDSS band magnitudes of image A and image B (see \citealt{Popovic2021}).
        }
\label{characteristics} 
\end{table*}

\section{Intrinsic variability and microlensing}\label{3}
\begin{figure*} 
\centering
\vspace*{3mm}
\hspace*{-3mm}
\includegraphics[width=18.15cm]{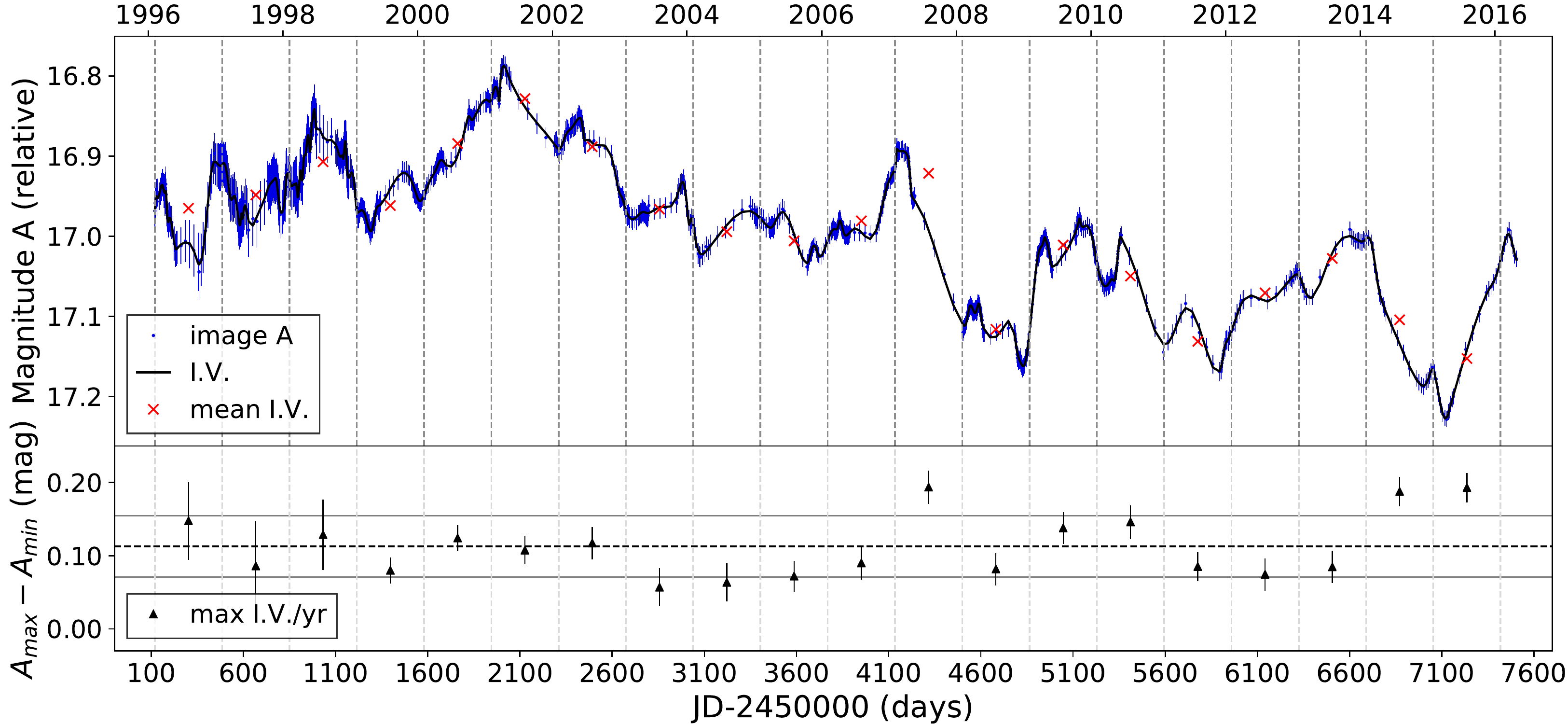}
\caption{Top: Intrinsic variability (spline fitting to the A light curve) plus mean variability per year (red crosses). Bottom: Maximum magnitude changes of the intrinsic variability per year.}
\label{IV}
\end{figure*}

Quasars are time variable, and since the images of multiple lensed quasars arrive with relative delays ranging from hours to several years because of the different paths taken by their light, variability of the source can mimic flux ratio anomalies. However, studies of optical continuum variability in gravitationally lensed quasars have the advantage that one is usually able to disentangle intrinsic from extrinsic variability (e.g., \citealt{Oscoz1996,Oscoz1997,Kundic1997,Paraficz2006,Goicoechea2008,Shalyapin2008}). To analyze the light curves of Q 0957+561 for the presence of extrinsic variations (i.e., microlensing), we first have to model and remove this variability that is intrinsic to the quasar itself. To accomplish that, we used the most recent time delay estimates (\citealt{Shalyapin2008}) and shifted the light curve of image B by -417 days. Thereafter, we corrected for the magnitude difference between the images using radio data from the literature (see Table \ref{radio}), assuming that these data represent the true magnification ratios of the images in the absence of microlensing. The radio-emitting regions of quasars should provide a realistic baseline for no microlensing, as they are supposed to arise from a large enough region to be insensitive to microlensing (see, e.g., \citealt{Metcalf2002,Mediavilla2009}). We assume that the flux variations in image A are mainly intrinsic as light from this component passes far from the lens galaxy.\footnotetext[1]{https://lweb.cfa.harvard.edu/castles/}
\footnotetext[2]{https://research.ast.cam.ac.uk/lensedquasars/}
\footnotetext[3]{https://esahubble.org/images/potw1403a/} Early studies (see, e.g., \citealt{Schild1991}) already attributed any differences in brightness between the A and B image to microlensing of the B component as the surface mass density of the lens galaxy is much lower at the position of component A. Less than $0.05\%$ (see \citealt{Jimenez2015a}) of the mass is expected to be in compact objects at this distance from the lens galaxy, coupled with an Einstein crossing time of $\sim$12.4 years (see \citealt{Mosquera2011}), which implies unlikely microlensing events for this quasar component. Light from the B component, however, passes through the lens galaxy (image B appears about 1\arcsec\ away from the center of G1; see Figure \ref{image}), and the probability of microlensing by the densely packed stars in the lens galaxy is relatively high (\citealt{Young1981,Schild1990,Schild1991}). As image A is expected to be unaffected by microlensing, it gives us an accurate history of the quasar's intrinsic brightness fluctuations. We obtain a source variability of $\sim$0.4 mag over the total duration of photometric monitoring. In addition, we averaged the individual measurements into one-year bins and examined the mean and maximum changes in brightness during this time (see Figure \ref{IV}). We obtain a mean variation of $\sim$0.1 mag/year and a maximum variation of $\sim$0.2 mag/year, respectively. \\

In this work, we studied two different cases: in the first case we performed a linear interpolation of image A's light curve to generate a set of photometric measurements at the same epochs of observations as those of the shifted light curve of image B. In the second case, we estimated the amplitude of the intrinsic variability by performing a single spline fitting to the A light curve (the image less prone to microlensing). Finally, we can subtract from the shifted light curve of image B the interpolated light curve/spline fitting of image A, creating a difference light curve (residuals) in which only the uncorrelated variability remains. \\

\begin{figure*} 
\centering
\includegraphics[width=18cm]{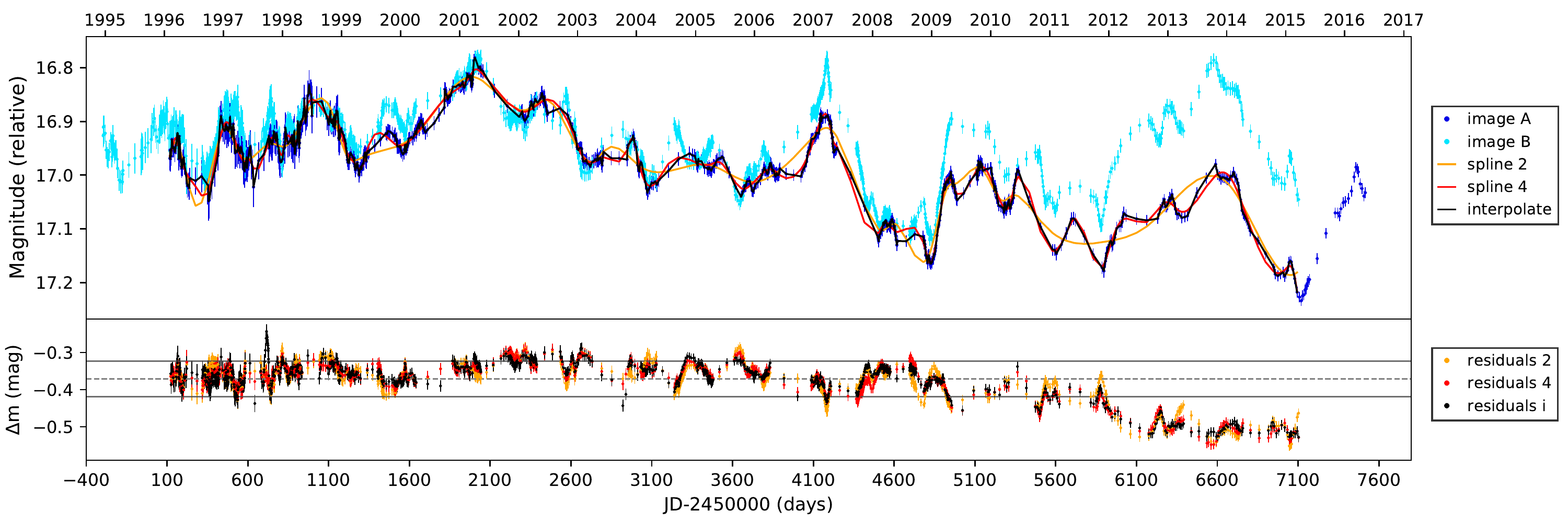}
\caption{Top: Image A and B light curves of Q 0957+561 in their overlapping region after shifting by the time delay. The linear interpolation of image A's light curve is shown in black, and two different models for the intrinsic variability of the quasar (splines with different knot steps fit to light curve A) are shown in red and in orange. To avoid confusion, we only display two out of four spline fittings. Bottom: Differential microlensing variability of the light curve B compared to the linear interpolation (black)/spline fits (red and orange) to light curve A. The dashed horizontal line shows the mean value of the residuals. We note that the residuals have been corrected for the magnitude difference between the images using radio data.}
\label{spline_residuals}
\end{figure*}

In Figure \ref{spline_residuals}, we present the time-delay shifted light curves of Q 0957+561 together with the residuals calculated from $\Delta m_B = m_B-m_{A(fit)}-(mB-mA)_{radio}$. From the upper panel in Figure \ref{spline_residuals}, we can see that during the first ten years (1996-2006) the light curves overlap well; then, a brightening of image B relative to image A is visible, which can be directly related to microlensing caused by stars in the lensing galaxy. Despite the similarity between the A and B light curves in this decade, some contribution of weak microlensing variability cannot be completely discarded. However, this is irrelevant because in our treatment we also consider the contributions of image A to microlensing in the simulated difference light curves. In the lower panel of Figure \ref{spline_residuals}, we can also clearly see the well-known quasi-constancy of the residuals of the B image between 1996 and 2006. In more recent years, higher variability took place (\citealt{Hainline2012,Shalyapin2012,Gil-Merino2018}) and the new results support the claim of \citealt{Hainline2012} that a microlensing event occurred during these years. The duration of this microlensing event (starting in 2009 and lasting at least until 2015) is still unclear and further future monitoring to map the full extent of the event will be needed. During this six-year range, the magnitude difference between the light curves increased from $\sim$0.1 mag to 0.2 mag. The low microlensing variability together with the long timescale are consistent with the relatively large optical quasar size of $\sim$12 lt-days reported in \citealt{Hainline2012}. 

\section{Bayesian source size estimation}\label{4}
We use a quantitative Bayesian method together with our determinations of the microlensing magnification amplitude to estimate the accretion disk size in the Q 0957+561 lensed quasar. The basic idea is to compare the histogram of microlensing magnification obtained from the observations (corresponding to the monitoring time interval) with the simulated predictions of microlensing variability for sources of different size (see \citealt{Fian2016,Fian2018}). 

\subsection{Simulated microlensing histograms}
To simulate the microlensing of extended sources, we use microlensing magnification maps created (for each quasar image) with the inverse polygon mapping method described in \citet{Mediavilla2006,Mediavilla2011}. Such maps show the microlensing magnification at a given source position and are determined by the local convergence, $\kappa$, and the local shear, $\gamma$, which can be obtained by fitting a singular isothermal sphere with an external shear (SIS+$\gamma_e$) to the coordinates of the images. The values of $\kappa$ and $\gamma$ for images A and B (taken from \citealt{Mediavilla2009}) are listed in Table \ref{lensmodel}. A magnitude difference of $m_{B-A} = -0.30$ mag was used for the macromodel, inferred from the average line-flux ratios of Mg II, C III], C IV, N V, and Ly$\alpha$ (taken from \citealt{Goicoechea2005}), consistent with the more recent estimate of the magnitude difference in Mg II reported by \citet{Motta2012}. We used a surface mass density in stars of $\alpha=10\%$ (\citealt{Mediavilla2009}) and generated 2000$\times$2000 pixel$^2$ magnification maps with a resolution of 0.2 lt-days per pixel (much smaller than the size of the optical accretion disk of the quasar), spanning 17.4$\times$17.4 Einstein radii$^2$. The value of the Einstein radius for this system is $3.25\times 10^{16} \sqrt{M/0.3M_\odot}$ cm $= 12.55 \sqrt{M/0.3M_\odot}$ lt-days at the source plane (\citealt{MosqueraKochanek2011}). We randomly distribute stars of a mass of $M = 0.3 M_\odot$ across the microlensing patterns to create a microlens convergence of 10$\%$. The ratio of the magnification in a pixel to the average magnification of the map gives the microlensing magnification at the pixel and histograms of normalized to the mean maps deliver the relative frequency of microlensing magnification amplitude of a pixel-size source.
\begin{table}[h]
\renewcommand{\arraystretch}{0.6}
\caption{Lens model parameters.}
\begin{tabu} to 0.49\textwidth {X[c]X[c]X[c]}
\hline 
\hline \\
\vspace*{-0.3cm}Image & \vspace*{-0.25cm}$\kappa$ & \vspace*{-0.25cm}$\gamma$ \\ 
\hline \\ 
\vspace*{-0.35cm}A & \vspace*{-0.35cm}0.20 & \vspace*{-0.35cm}0.15 \\ 
\vspace*{-0.25cm}B & \vspace*{-0.25cm}1.03 & \vspace*{-0.25cm}0.91 \\ \hline 
\end{tabu}
\label{lensmodel}    
\end{table}

To model the structure of the unresolved quasar source, we considered a circular Gaussian intensity profile of size $r_s$, $I(R) \propto exp(-R^2/2r_s^2)$. It is generally accepted that the specific shape of the source's emission profile is not important for microlensing flux variability studies, since the results are mainly controlled by the half-light radius rather than by the detailed source profile (\citealt{Mortonson2005}). The characteristic size $r_s$ is related to the half-light radius, that is, the radius at which half of the light at a given wavelength is emitted, by $R_{1/2} = 1.18 r_s$. As lengths are measured in Einstein radii, which scale as $R_E \propto \sqrt{M}$, all calculated sizes can be rescaled accordingly for a different mean stellar mass. Finally, we convolve the magnification maps with Gaussians of 21 different sizes over a linear grid that spans from $r_s = 0.5$ to 40.5 lt-days and after convolution, we normalize each map by its mean value. The histograms of the normalized map represent the histograms of the expected microlensing variability. Thus, we obtain 21 different microlensing histograms corresponding to different source sizes. The movement of a extended source across the magnification map (appearing as a network of high-magnification caustics separated by regions of lower magnifications) is equivalent to a point source moving across a map that has been smoothed by convolution with the intensity profile of the source. Large values or $r_s$ smear out the network of microlensing magnification caustics and reduce their dynamic range, thereby causing the histograms to become narrower. Finally, convolving the histograms of B with the histogram of A, we build the microlensing difference histograms B-A for different values of $r_s$ to be compared with the observational histograms obtained from the light curves (see Section \ref{42}). We adopt a bin size of 0.05 mag for the modeled and experimental microlensing histograms.

\subsection{Observed microlensing histograms}\label{42}
\begin{figure}[h] 
\centering
\includegraphics[width=9cm]{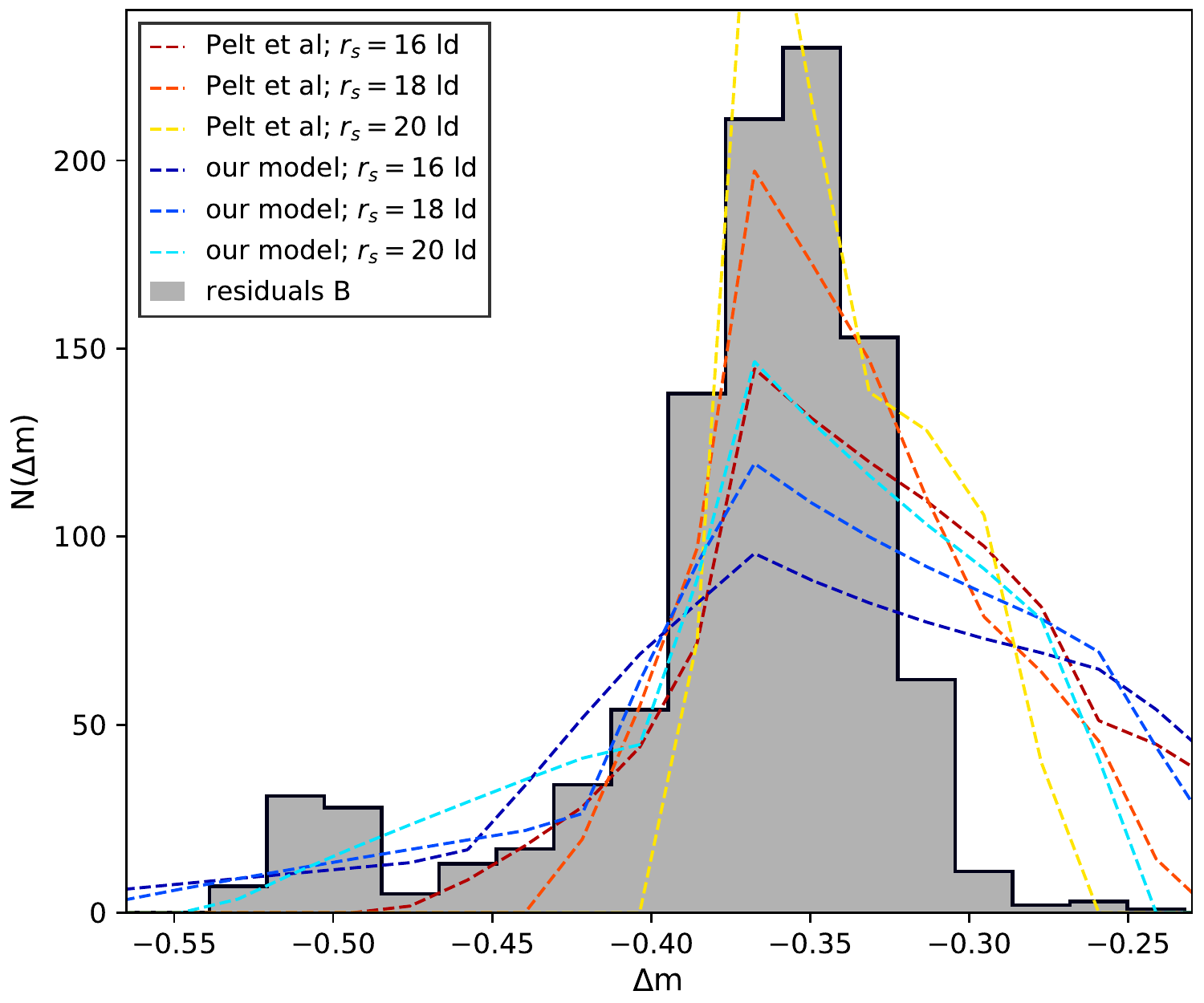}
\caption{Microlensing frequency distributions obtained from the observations, i.e, the difference light curve presented in the lower panel of Figure \ref{spline_residuals} (gray histogram), and the simulated magnification maps (segmented lines). The various segmented lines correspond to convolutions of the simulated 
magnification maps with sources of different sizes (in \textit{\emph{blue}} shades for our macrolens model and in \emph{\textit{\emph{red}}  }shades for the model of \citealt{Pelt1998}).}
\label{model_hist}
\end{figure}

The effect of a finite source size is that it smooths out the flux variations in the light curves of lensed quasars caused by stars in the galaxy. From the residual light curves that represent the differential (with respect to A, the image less prone to microlensing) microlensing of the B image, we obtained the microlensing variability histogram; this refers to the frequencies at which each microlensing amplitude appears in the microlensing variability light curves. In Figure \ref{model_hist}, we compare the B-A modeled magnification histograms corresponding to convolutions with sources of different sizes (dashed lines) with the experimental microlensing histograms.

\subsection{Method}
To study the likelihood of different values adopted for the size, we compare the microlensing histograms inferred from the model (corresponding to the convolutions with different source sizes) with the histogram of the data using the following statistic:
\begin{equation}
P_X(r_s) = \sum_{i=1}^{N_{bin}} h_{X-B}^{i}\ \hat{h}_{X-B}^{i}(r_s),
\end{equation}
where $h_{X-B}^{i}$ and $\hat{h}_{X-B}^{i}(r_s)$ are the observed and modeled histograms, and $N_{bin}$ is the number of bins. This histogram product is based on the distance between histograms (related to the Pearson correlation coefficient) and is a natural extension of the single-epoch method.

\section{Accretion disk size and impact of uncertainties}\label{5}
To check the robustness of our microlensing based method used to estimate the size of the quasar's accretion disk, we study the impact of different uncertainties and sources of systematic errors.

\subsection{Uncertainties in fitting the intrinsic variability}\label{sIV}
To study the importance of accurately fitting the intrinsic variability, we studied five different cases. In the first case, we performed a linear interpolation of image A's light curve and subtracted it from the shifted B light curve (see Figure \ref{spline_residuals}). Given the available data with high signal–to–noise ratio (S/N), we are able to produce a reasonable set of residuals (see lower panel in Figure \ref{spline_residuals}). In the other four cases, the idea was to fit one single spline representing the intrinsic variation of the quasar. For each spline, we used a different smoothness (or flexibility), meaning we changed the initial spacings (in days) of the knots and repeated the same procedure as in the first case. From Figure \ref{spline_residuals}, we see that although the splines with the lowest knot steps (e.g., spline 1 and spline 2) cannot adequately capture the underlying structure of the data and do not fit fast changes in magnitude well, the resulting residuals look almost the same. Hence, neither underfitting nor overfitting produce significant changes in the residuals for light curves with moderate SNR. Producing histograms of the different sets of residuals and comparing them with the simulated histograms for different sizes of $r_s$, we obtain very similar results in all five cases (see Table \ref{IVmodel}).

\begin{table}[h]
\renewcommand{\arraystretch}{1.4}
\caption{Accretion disk size using different models for the quasar's intrinsic variability.}
\begin{tabu} to 0.49\textwidth {X[l]X[c]X[c]} 

\hline
\hline 
I.V. Model & $r_s$ (lt-days) & $R_{1/2}$ (lt-days) \\ 
\hline 
Interpolation & $14.9_{-10.4}^{+7.6}$ & $17.6_{-12.3}^{+9.0}$ \\ 
Spline 1 & $14.8_{-10.3}^{+7.7}$ & $17.5_{-12.2}^{+9.1}$ \\ 
Spline 2 & $14.8_{-10.3}^{+7.7}$ & $17.5_{-12.2}^{+9.1}$ \\
Spline 3 & $14.9_{-10.4}^{+7.6}$ & $17.6_{-12.3}^{+9.0}$ \\
Spline 4 & $14.9_{-10.4}^{+7.6}$ & $17.6_{-12.3}^{+9.0}$ \\ \hline 
\end{tabu}
\label{IVmodel}    
\end{table}

\subsection{Uncertainties in the time delay}\label{std}
Intrinsic variations in brightness records of gravitationally lensed quasars can be used to estimate global time delays between components (\citealt{Ref1964}), and after several years of analyzing different sets of data from various telescopes, the scientific community appears to be converging on a time-delay value near 400 days (see Table \ref{timedelay}). A serious problem for the time-delay estimation from the optical monitoring data had been the imperfect sampling, since brightness changes on time scales of days and weeks have been observed (\citealt{Schild1997}). We check the influence of uncertainties in the time delay on the size estimates by adopting a value of $\Delta t=417$ days (taken from \citealt{Shalyapin2008}) and additionally shifting the light curves by $\pm$1$\sigma$ and $\pm$2$\sigma,$ respectively. The relatively small uncertainties in the time delay (of the order of a few days) do not induce significant changes in the disk size measurement.

%\begin{table}[h]
%%\tabcolsep=0.3cm
%\renewcommand{\arraystretch}{1.4}
%\caption{Accretion Disk Size using different Time Delays.}
%\begin{tabu} to 0.49\textwidth {X[l]X[l]X[l]} 
%\hline 
%\hline 
%$\Delta t$ (days) & $r_s$ (lt-days) & $R_{1/2}$ (lt-days) \\ 
%\hline 
%409 & $14.8_{-10.3}^{+7.7}$ & $17.5_{-12.2}^{+9.1}$ \\ 
%411 & $14.8_{-10.3}^{+7.7}$ & $17.5_{-12.2}^{+9.1}$ \\ 
%413 & $14.8_{-10.3}^{+7.7}$ & $17.5_{-12.2}^{+9.1}$ \\
%415 & $14.8_{-10.3}^{+7.7}$ & $17.5_{-12.2}^{+9.1}$ \\
%417 & $14.8_{-10.3}^{+7.7}$ & $17.5_{-12.2}^{+9.1}$ \\ \hline 
%\end{tabu}
%\label{dt}    
%\end{table}

\subsection{Uncertainties in the baseline for no microlensing}
We studied the change of the accretion disk size when using different radio and emission line flux ratio measurements from the literature (see Table \ref{radio}). In a spectroscopic analysis, \citet{Motta2012} used broad emission lines (BELs) to estimate the base for no microlensing, finding that the B-A magnitude differences follow a decreasing trend toward the blue compatible with extinction. Toward the red, the flux ratios are fully consistent with the B-A radio measurements (which are uncontaminated by the lens galaxy continuum and free from dust extinction). We found that the size measurements are sensitive to the baseline for no microlensing, in the sense that a flux ratio change of 0.1 results in a difference of $\sim$5 lt-days (see Table \ref{radiosize}). However, the uncertainties in the latest radio measurements are relatively small ($\sim$0.02; see \citealt{Haarsma1999}), leading to a difference of only one lt-day.
\begin{table}[h] 
\tabcolsep=0.36cm
\renewcommand{\arraystretch}{1.2}
\caption{Radio and BEL measurements from the literature.}
\begin{tabu} to 0.49\textwidth {lll}

\hline
\hline 
B/A & $\lambda$ & Reference \\ 
\hline 
 0.82$\pm$0.02 & 13 cm & \citealt{Gorenstein1988} \\ 
 0.79$\pm$0.05 & 18 cm& \citealt{Garrett1990} \\ 
 0.752$\pm$0.028 & 6 cm & \citealt{Conner1992} \\
 0.723$\pm$0.044 & 6 cm & \citealt{Conner1992} \\
 0.76$\pm$0.03 & 13 cm & \citealt{Chartas1995} \\
 0.75$\pm$0.02 & 18 cm & \citealt{Garrett1994} \\
 0.74$\pm$0.02 & 4 and 6 cm & \citealt{Haarsma1999} \\ 
 0.759$\pm$0.007 & Mg II (MMT) & \citealt{Motta2012} \\
 0.787$\pm$0.022 & Mg II (HST) & \citealt{Motta2012} \\
 $\sim$0.69 & several BELs & \citealt{Motta2012} \\ \hline 
\end{tabu}
\label{radio}    
\end{table}

\begin{table}[h] 
\renewcommand{\arraystretch}{1.4}
\caption{Accretion disk size using different radio and BEL ratios.}
\begin{tabu} to 0.49\textwidth {X[l]X[l]X[l]} 

\hline
\hline 
$B/A$ & $r_s$ (lt-days) & $R_{1/2}$ (lt-days) \\ 
\hline 
\hspace*{-2pt} 0.69 & \hspace*{-2pt} $13.1_{-8.6}^{+7.4}$ & \hspace*{-2pt} $15.5_{-10.7}^{+8.7}$ \\
0.72 & $14.1_{-9.6}^{+8.4}$ & $16.6_{-11.3}^{+10.1}$ \\ 
0.73 & $14.5_{-10.0}^{+8.0}$ & $17.1_{-11.8}^{+9.4}$ \\ 
0.74 & $14.8_{-10.3}^{+7.7}$ & $17.5_{-12.2}^{+9.1}$ \\
0.75 & $15.4_{-10.9}^{+9.1}$ & $18.2_{-12.9}^{+10.7}$ \\
0.76 & $15.8_{-9.3}^{+8.7}$ & $18.6_{-11.0}^{+10.3}$ \\ 
0.77 & $16.4_{-11.9}^{+10.1}$ & $19.4_{-14.0}^{+11.9}$ \\
0.78 & $16.8_{-12.3}^{+9.7}$ & $19.8_{-14.5}^{+11.4}$ \\
0.79 & $17.1_{-12.6}^{+11.4}$ & $20.2_{-14.9}^{+13.5}$ \\
0.80 & $17.9_{-13.4}^{+10.6}$ & $21.1_{-15.8}^{+12.5}$ \\
0.81 & $18.3_{-11.8}^{+12.2}$ & $21.6_{-13.9}^{+14.4}$ \\
0.82 & $18.8_{-12.3}^{+11.7}$ & $22.2_{-14.5}^{+13.8}$ \\ \hline 
\end{tabu}
\label{radiosize}    
\end{table}

\subsection{Uncertainties in the lens model}
The values of the convergence, $\kappa$, and shear, $\gamma$, at the location of each image are needed to compute the magnification maps from which the simulated microlensing histograms are obtained. These parameters are inferred from the macrolens model and may be affected by uncertainties. We checked for the robustness of our results with respect to the macromodel by comparing it with the lens parameters from \citealt{Pelt1998} (see Table \ref{macropelt}). After computing the magnifications maps for images A and B, we repeat all of the calculations, obtaining a slightly smaller value for the half-light radius of the accretion disk ($R_{1/2} = 12.5_{-7.2}^{+9.3}$ lt-days). The relationship of the errors in the macromodel with the uncertainty in the disk size is not simple and probably depends on the change in the magnification of the source.

\begin{table}[h] 
%\centering
\tabcolsep=0.53cm
\renewcommand{\arraystretch}{1.3}
\caption{Lens model parameters from \citealt{Pelt1998}.}
\begin{tabu} to 0.49\textwidth {X[c]X[c]X[c]X[c]X[c]} 

\hline
\hline 
Image & $\kappa$ & $\gamma$ & $\Delta \kappa^*$ & $\Delta \gamma^*$ \\
\hline 
A & 0.22 & 0.17 & 0.02 & 0.02 \\
B & 1.24 & 0.90 & 0.21 & 0.01 \\
\hline 
\end{tabu}
\label{macropelt}
\tablefoottext{*}{difference compared to our model}
\end{table}

\subsection{Uncertainties in fraction of mass in stars}
The amplitude of microlensing is sensitive to the local stellar-surface mass-density fraction (as compared to that of the dark matter at the image location (see, e.g.,  \citealt{Schechter2002})). Hence, the source size is sensitive to the stellar mass fraction $\alpha$. In Q 0957+561, the two lensed images are located at very different radii from the center of the lens, resulting in different fractions of convergence in the form of stars. \citet{Jimenez2015a} measured the stellar mass fraction at two different radii from a sample of 18 different lens system with available $R_E/R_{eff}$, where $R_E$ is the Einstein radius and $R_{eff}$ is the effective radius of the lens galaxy. They found that the stellar-surface mass density is $\alpha\sim0.3$ at a radius of ($1.3\pm0.3$) $R_{eff}$. We recomputed the magnification maps using a value $\alpha=0.3$ for image B ($R_E/R_{eff} = 1.29$ for Q 0957-561; see \citealt{Fadely2010}), and adopting a value of $\alpha=0$ for image A, which is equivalent to not using any magnification map for this image (due to the contribution of the cluster gravitational potential, this image is located far from the lens galaxy). We repeated all the calculations and obtained a slightly bigger value for the half-light radius of the accretion disk ($R_{1/2} = 21_{-13.9}^{+14.4}$ lt-days). When $\alpha=0$ for image A, the 
B-A microlensing magnification probability can be directly inferred 
from the histogram of microlensing magnifications corresponding to 
image B. As a cross-check, we repeated the calculations considering only the magnification map for image B, obtaining the same result.

\subsection{Probability density function of the source size}
\begin{figure} 
\includegraphics[width=8.86cm]{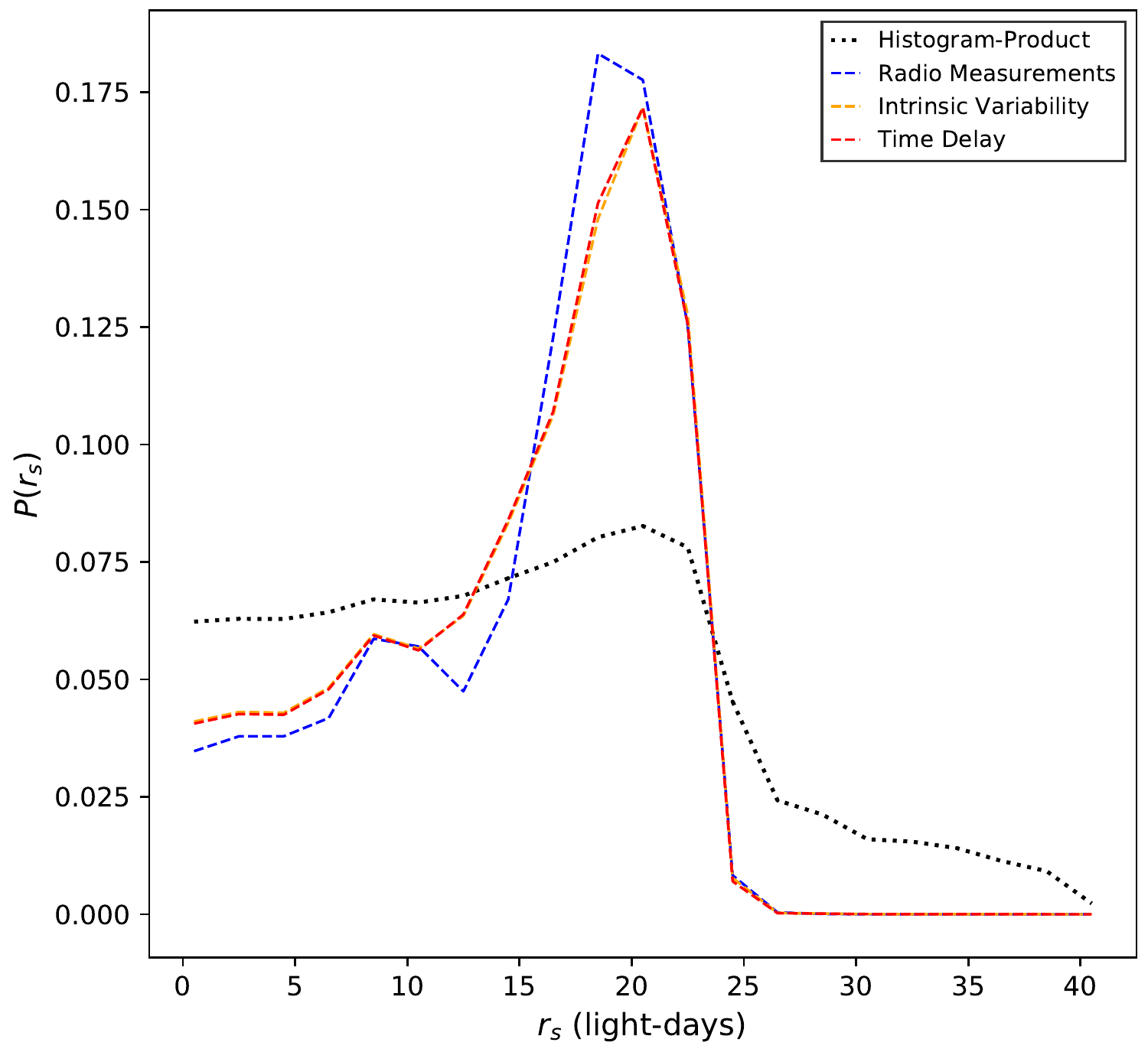}
\caption{Probability density distribution of the source size $r_s$ using a time delay of $\Delta t = 417$ days, a radio-flux ratio of $B/A= 0.74$, a linear interpolation of image A's light curve as an intrinsic variability model, and the lens model parameters of \citet{Mediavilla2009} (dotted \textit{black} line). The dashed lines show the PDFs for various model/data-related analyses as indicated by the legend. These are obtained by multiplying single probability distributions corresponding to different time delays (\textit{red}), models for the intrinsic variability (\textit{orange}; see Table \ref{IVmodel}), and radio measurements (\textit{blue}; see Table \ref{radiosize}). We note that the \textit{red} and \textit{orange} distributions overlap.}
%\caption{Probability distributions of the source size $r_s$ for various models/data-related analysis (dashed lines), as indicated by the legend. The thick black line represents the product of all the distributions, which accounts for the different systematic and model dependent uncertainties.}
\label{PDFs}
\end{figure}

To evaluate the impact that changes in the previously discussed values/models can have on the results, we multiplied all the possible probability distributions for each source of uncertainty (see Figure \ref{PDFs}). That means, in the case of the intrinsic variability, we multiplied five probability distributions; for the radio/BEL measurements, we multiplied twelve probability distributions; and so on. In Table \ref{size}, we summarize the contributions of the different sources of uncertainty on the size estimates and list their deviation from the inferred disk size when using the most recent values for both, the time delay ($\Delta t =417$ days), and the radio flux ratio ($B/A =0.74$), a linear interpolation for the image's A light curve as an intrinsic model, and the lens model parameters of \citet{Mediavilla2009}. The uncertainties in modeling the intrinsic variability as well as the uncertainties in the time delay have no significant influence on the size. With regard to the baseline of no microlensing, assuming very conservative errors, we obtain an increase of $\sim$3\% in the estimate of the size. Thus, we get $\sim$30\% smaller sizes for the accretion disk using the lens model of \citealt{Pelt1998}, and $\sim$20\% bigger sizes when using a different fraction of mass in stars for both images. Finally, we obtain a half-light radius of $R_{1/2} = 17.6\pm6.1\sqrt{M/0.3M_\odot}$ lt-days for the region emitting the R-band emission using $\sqrt{\sum_{i} {\Delta R_{1/2}}_i^2}$ from Table \ref{size} to estimate the uncertainty in size (i.e., the 1$\sigma$ variation of the peak of the probability distributions in each direction). %The resulting probability distribution can be seen in Figure \ref{PDFs}. 

\begin{table}[h]
\renewcommand{\arraystretch}{1.4}
\caption{Impact of uncertainties on the size estimates.}
\begin{tabu} to 0.49\textwidth {X[l]X[c]X[c]X[c]} 

\hline
\hline 
Source & $\ \ r_s\ \ $ (lt-days) & $R_{1/2}$ (lt-days) & $\ \ \ \ \Delta R_{1/2}^{\ a}\ \ \ $  (lt-days)\\ 
\hline 
I.V.$^{\ b}$ & $14.8_{-10.3}^{+7.7}$ & $17.5_{-12.2}^{+9.1}$ & $-0.1$ \\ 
$\Delta$t & $14.8_{-10.3}^{+7.7}$ & $17.5_{-12.2}^{+9.1}$ & $-0.1$ \\ 
\hspace*{-2pt}Baseline & \hspace*{-2pt}$15.3_{-6.8}^{+5.1}$ & \hspace*{-2pt}$18.1_{-8.0}^{+6.0}$ & \hspace*{-2pt}$+0.4$ \\
Model & $10.6_{-6.1}^{+7.9}$ & $12.5_{-7.2}^{+9.3}$ & $-5.1$ \\
\hspace*{-2pt}$\alpha^{\ c}$ &\hspace*{-2pt} $18.3_{-11.8}^{+12.2}$ & \hspace*{-2pt} $21.0_{-13.9}^{+14.4}$ &\hspace*{-2pt} $+3.4$ \\ \hline 
\end{tabu}
\label{size}  
\tablefoottext{a}{Deviation from the inferred disk size of $R_{1/2}=17.6$ lt-days\\ \hspace*{3mm} (see dotted black line in Figure \ref{PDFs}).}\\
\tablefoottext{b}{Intrinsic variability}\\
\tablefoottext{c}{Mass in stars}
\end{table}

\subsection{Impact of the effective velocity on the size estimate}
\begin{figure} 
\includegraphics[width=9.cm]{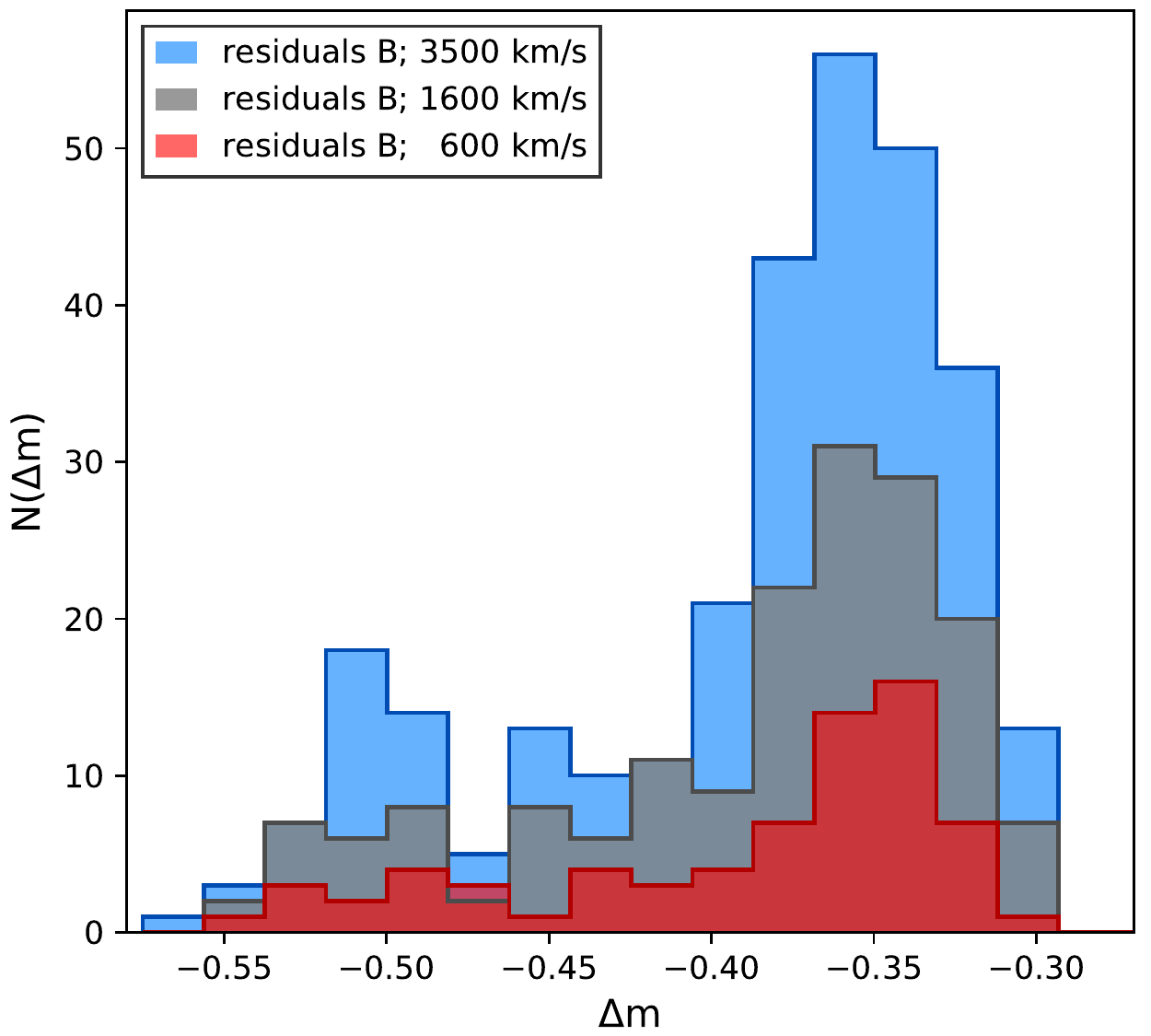}
\caption{Microlensing frequency distributions obtained from the (according to the effective velocity averaged) difference light curves for $\nu_e = 1600\ km\ s^{-1}$ (\textit{gray}), $\nu_e = 600\ km\ s^{-1}$ (\textit{red}), and $\nu_e = 3500\ km\ s^{-1}$ (\textit{blue}).}
\label{velocity}
\end{figure}

The information on the observed light curve can be better extracted by full fitting procedures (\citealt{Kochanek2004,Hainline2012, Cornachione2020}). This procedure is beyond the scope of this paper, yet it is not the only possible approach to use the observed light curve. Even if we do not use the time-ordered series of data, we have presented a simple, fast statistical approach to extract information on the source size from the light curve (e.g., \citealt{Fian2016}). Nevertheless, the effective velocity of the source can still affect the timescale on which points in the observed light curves have to be compared with our models. To take this effect into account, we tested the influence of the velocity on our size estimate by sampling (i.e., averaging using a certain window) the experimental light curves with the time window corresponding to the magnification map pixel size and the effective velocity for the source trajectory. \citet{Hainline2012} constructed the effective source plane (Einstein) velocity, $\nu_e$, using the method described in \citet{Kochanek2004}, applying the measured stellar velocity dispersion for the lens galaxy ($\sigma_\star = 288 \pm 9$ km s$^{-1}$; \citealt{Tonry1999}), and obtaining the dispersion of the peculiar velocity distribution at the redshifts of the quasar and the lens galaxy from the power-law fits by \citet{Mosquera2011} to the peculiar velocity models of \citet{Tinker2012}. As a consequence of the low intrinsic variability of Q 0957+561, coupled with the relatively low amplitude of the microlensing signal, they obtained a wide velocity distribution with a median of 1600 km s$^{-1}$ and a 68\% confidence range of $600\ km\ s^{-1} < \nu_e < 3500\ km\ s^{-1}$. The median velocity corresponds to a time window of 1.2 months for a pixel size of 0.2 lt-days ($\sim$ 200 points in the 21 years long light curves). The $+1 \sigma$ velocity coincides with a time window of 0.54 months (i.e., $\sim$ 480 data points), and the $-1\sigma$ velocity with a time window of 3.3 months ($\sim$ 70 data points). We built three experimental histograms (see Figure \ref{velocity}) of the (according to the velocity averaged) microlensing difference light curves and compared them with the simulated histograms for different sizes of $r_s,$ as described in Section \ref{4}. We found that the overall shape of the experimental histograms stays the same and that the velocity does not induce significant changes in the size estimates (with a maximum change of $\sim$0.1 lt-days).

\section{Conclusions}\label{6}
We studied 21 years of monitoring data for the lensed images of the twin quasar Q 0957+561, which so far are the longest available light curves of a gravitational lens system. Unlike most other known lens systems, photometric monitoring of this object is relatively easy since the system is relatively bright (I=16) and because of its wide image separation ($\sim$6''). We used the GLENDAMA light curves (\citealt{Gil-Merino2018}) to obtain the accretion disk size, which with more than thousand epochs of observation significantly extend the time coverage of previous works. Taking image A, which is less affected by microlensing, as a reference and using the experimental time delays inferred by \citealt{Shalyapin2008}, we removed the intrinsic variability from light curve B in the overlapping region. Using the radio-flux ratio between components A and B measured by \citealt{Haarsma1999} as a baseline for no microlensing, we finally obtained the microlensing light curve B-A. We found microlensing of up to 0.5 mag in the residuals of recent years and we used the statistics of microlensing during all available seasons to infer probabilistic distributions for the source size. Using the product of the observed and modeled microlensing histograms, we obtained a half-light radius of $R_{1/2} = 17.6\pm6.1\sqrt{M/0.3M_\odot}$ lt-days. According to Table \ref{size}, uncertainties in the faction of mass in stars and in the macromodel are the dominant sources of error. However, uncertainties corresponding to these values in the error budget are probably conservative upper limits. In Section \ref{sIV}, we see that there is no significant dependence on the model used to simulate the quasar's intrinsic variability. Thus, small changes in the estimation of the time delay between images have no effect on the source size (see Section \ref{std}). This supports both the robustness of the method and the limited impact of uncertainties on the accretion disk size. \\

Our estimate for the size is significantly greater than the average determinations obtained for a sample of lensed quasars by \citet{Jimenez2012,Jimenez2014}\ ($R_{1/2} (2558\AA) \sim 8$ lt-days and $R_{1/2} (2558\AA) \sim 10$ lt-days, respectively) when scaling the rest wavelength (from 1736\AA\ and 1026\AA, respectively) to 2558\AA\ (using $R_{1/2} = (\lambda_0/\lambda)^p\ R_{1/2}(\lambda)$) for microlenses with a mean mass of $M = 0.3 M_\odot$ and assuming $\alpha=10\%$. Comparing with the average accretion disk size obtained by \citet{Jimenez2015b} when using a more realistic value for the fraction of mass in stars ($\alpha=20\%$), their reported values of $\sim$10 lt-days (maximum likelihood) and $\sim$8 lt-days (Bayesian) at $\lambda_{rest}=1736$\AA\ correspond to $\sim$16 lt-days and $\sim$13 lt-days at 2558\AA\ (for microlenses with a mean mass of $M = 0.3 M_\odot$), matching our estimate for the disk size in Q 0957+561. Our result appears to be consistent within errors with the R-band half-light radius of \citealt{Hainline2012} ($R_{1/2} = 12.2_{-8.3}^{+26.4}$ lt-days). Furthermore, our results are in excellent agreement with the continuum size recently predicted by \citealt{Cornachione2020} ($R_{1/2} = 17.6_{-13.4}^{+23.8}$ at $\lambda_{rest}=2558\AA$) using the light curve fitting method (see \citealt{Kochanek2004}). Thus, we observe that the microlensing size is significantly larger than the prediction of the thin disk theory, also found for other lensed quasars by \citet{Pooley2007}, \citet{Morgan2010}, and \citet{Blackburne2011}. We note that the broad-line region could contribute substantially to the continuum level around the Mg II line due to the underlying iron blends, the Balmer recombination edge, and the Mg II line itself. The degree to which recent findings for low-luminosity sources are also relevant for high-luminosity quasars is still unclear (see, e.g., \citealt{Chelouche2019,Korista2001,Korista2019,Lawther2018}).\\

%Another possibility to explain the relatively large source size obtained for this system is a low value of the effective transverse velocity. This velocity describes how fast the source is moving through the magnification pattern and, consequently, if the quasar image is in a region of low microlensing magnification and the effective transverse velocity is small, we can observe a low amplitude of microlensing magnification, independently Iof the source size. Thus, the observed long-timescale microlensing induced flux variations in Q 0957+561 support the theory of a very small effective velocity with respect to the magnification pattern. We leave an elaborated discussion of the introduction of the velocity in the Bayesian analysis for a forthcoming paper.\\

\begin{acknowledgements}
We thank the anonymous referee for the helpful comments, and constructive remarks on this manuscript. We thank the GLENDAMA project for making publicly available the monitoring data of Q 0957+561. C.F. gratefully acknowledges the financial support from Tel Aviv University and University of Haifa through a DFG grant HA3555-14/1. E.M. and J.A.M  are supported by the Spanish MINECO with the grants AYA2016- 79104-C3-1-P and AYA2016-79104-C3-3-P. J.A.M. is also supported from the Generalitat Valenciana project of excellence Prometeo/2020/085. J.J.V. is supported by the project AYA2017-84897-P financed by the Spanish Ministerio de Econom\'\i a y Competividad and by the Fondo Europeo de Desarrollo Regional (FEDER), and by project FQM-108 financed by Junta de Andaluc\'\i a. K. R. acknowledges support from the Swiss National Science Foundation (SNSF). V.M. acknowledges the support of Centro de Astrof\'{\i}sica de Valpara\'{\i}so.  
\end{acknowledgements}

\bibliographystyle{aa}
\bibliography{bib}

\end{document}